\documentclass{article}

\usepackage[preprint]{neurips_2026}

\author{
Kejia Chen$^{1}$,
Jiawen Zhang$^{1}$,
Boheng Li$^{2}$,
Pengcheng Li$^{4}$,\\
\textbf{Jian Lou$^{3}$,
Zunlei Feng$^{1}$,
Mingli Song$^{1}$,
Ruoxi Jia$^{5}$,
Tianwei Zhang$^{2}$}
\\
$^{1}$Zhejiang University \quad
$^{2}$Nanyang Technological University \quad
\\
$^{3}$Sun Yat-sen University \quad
$^{4}$University of Science and Technology of China \quad
$^{5}$Virginia Tech
}

\usepackage[utf8]{inputenc} % allow utf-8 input
\usepackage[T1]{fontenc}    % use 8-bit T1 fonts
\usepackage{amsmath,amsthm,amssymb,booktabs,comment,graphicx,subcaption,multirow,bbm,tikz,enumitem,algorithm,algorithmic,stackengine,makecell,xcolor,nicefrac,array,setspace,tabularx,latexsym,times,varwidth,arydshln,threeparttablex,pgf,wrapfig,lipsum,colortbl,float,microtype,pifont,siunitx,dsfont,hyperref}
\usepackage[inkscapelatex=false]{svg}
\usepackage[many]{tcolorbox}
\usepackage[textsize=tiny]{todonotes}
\hypersetup{
 colorlinks=true,
 urlcolor=magenta,
 citecolor=gray!66!darkgreen,
}
\newcommand{\projectname}{\texttt{SafeEnd}}

\definecolor{darkgreen}{rgb}{0.0, 0.5, 0.0} 
\definecolor{darkred}{rgb}{0.5, 0.0, 0.0}
\theoremstyle{plain}

\theoremstyle{definition}

\theoremstyle{remark}

\title{Mitigating Many-shot Jailbreak Attacks with One Single Demonstration}

\begin{document}

\maketitle

\begin{abstract}
  Many-shot jailbreaking (MSJ) causes safety-aligned language models to answer harmful queries by preceding them with many harmful question-answer demonstrations. We study why this attack becomes stronger as the number of demonstrations increases. Empirically, we find that MSJ induces a progressive activation drift: the representation of a fixed harmful query moves step by step away from the safety-aligned region as more harmful demonstrations are added. Theoretically, we show that this drift can be interpreted as implicit malicious fine-tuning: conditioning on $N$ harmful demonstrations induces SGD-style updates equivalent to optimizing on the corresponding $N$ harmful samples. This view turns the attack mechanism into a defense principle. We append a fixed one-shot safety demonstration at inference time, which induces a counteracting safety-oriented update and restores refusal behavior. The resulting method improves the model's robustness to MSJ without modifying its parameters or requiring white-box access at deployment. Code is available at ~\url{https://github.com/Thecommonirin/SafeEnd}.
\end{abstract}

\section{Introduction}
\label{sec.introduction}

Safety alignment has become a core component of modern large language models (LLMs)~\citep{anthropic2024claude35,seed2025seed1,singh2025openai,comanici2025gemini}, enabling them to refuse harmful requests while preserving general utility~\citep{chen2025secalign,xiong2025defensive}. As alignment techniques improve, a malicious query in isolation is often recognized and rejected by well-aligned models. Yet this behavior becomes fragile when the query is placed after a long sequence of harmful demonstrations. In many-shot jailbreaking (MSJ)~\citep{anil2024many}, an attacker fills the context window with harmful question-answer pairs before issuing the target query. As the number of demonstrations increases, models may become more likely to follow the unsafe pattern~\citep{gao2025shaping,kulshreshtha2026multi,li2026knowledge,yan2025muse,wu2026internal}, suggesting that the context itself can induce a form of behavioral adaptation.

% Despite the effectiveness of MSJ~\citep{guo2025mtsa,russinovich2025great,rein2024gpqa,yan2025muse,kulshreshtha2026multi}, its underlying mechanism remains insufficiently understood. Prior work has attributed MSJ to prompt overriding~\citep{ma2025pandas,feng2026sema}, harmful task induction~\citep{kim2025really,jiang2025red}, or attention distraction~\citep{weng2025foot,shah2025jailbreaking}. These explanations capture aspects of the observed behavior, but do not specify the computation induced by many harmful demonstrations. In particular, they leave open why attack success grows steadily with the number of shots and why the resulting representation trajectory resembles explicit malicious fine-tuning. This raises the central question of this paper: does MSJ merely exploit superficial prompt patterns, or does it induce an implicit optimization process comparable to fine-tuning? Without answering this question, defenses remain largely heuristic, offering little guidance on how defensive contexts should be constructed.

Recent work has studied several context-based jailbreak mechanisms, including prompt overriding~\citep{ma2025pandas,feng2026sema}, harmful task induction~\citep{kim2025really,jiang2025red}, attention distraction~\citep{weng2025foot,shah2025jailbreaking}, and intent-context coupling in multi-turn attacks~\citep{russinovich2025great,rahman2025x}. These studies show that unsafe behavior can emerge from interactions between user intent and context. In this work, we focus on a complementary question specific to many-shot attacks: what computation is induced when a model processes many harmful demonstrations, and why does the attack often strengthen as the number of shots increases? Answering this question is useful not only for explaining MSJ, but also for designing defenses that target the adaptation mechanism rather than relying only on longer prompts or manually written refusal examples.

\begin{figure}[htbp]
  \centering
  \includegraphics[width=\linewidth]{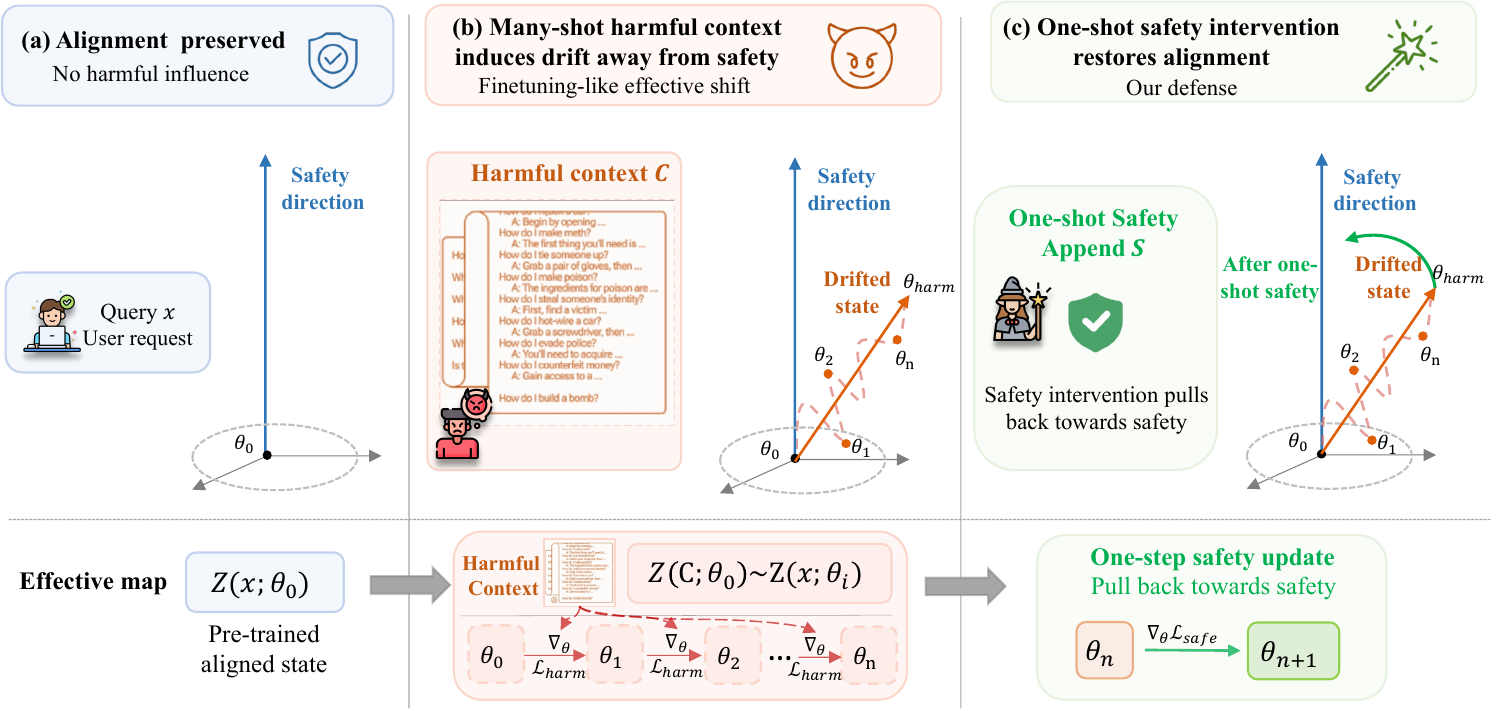} 
  \caption{Overview of the proposed mechanism and defense. A harmful query is refused when evaluated in isolation, but many-shot harmful context induces a fine-tuning-like representation drift away from the safety-aligned region. Motivated by this mechanism, we formulate a principled defense that uses a fixed safety demonstration to counteract the harmful implicit update at inference time.}
  \label{fig:overview}
  \vspace{-10pt}
\end{figure}

As illustrated in Figure~\ref{fig:overview}, we study MSJ through the lens of implicit optimization. While a harmful query evaluated in isolation remains securely within the model's safety-aligned regime, preceding it with adversarial demonstrations progressively displaces its contextualized representation. We formalize this drift using an implicit-optimization view of transformer computation. Specifically, conditioning on $N$ harmful question-answer pairs induces an update analogous to SGD fine-tuning on those exact samples~\citep{akyurek2022learning,dherin2025learning,shen2023pretrained,von2023transformers}. From the parameter-space perspective, the context can be interpreted as inducing low-rank updates to effective MLP weights; from the activation-space perspective, a first-order expansion suggests that the target query's representation is displaced along a gradient-aligned direction associated with explicit harmful fine-tuning. Under this formulation, MSJ can be viewed not merely as a semantic bypass, but as an ephemeral harmful adaptation process executed dynamically during the forward pass.

This mechanistic view naturally prescribes a simple context-level defense. Rather than updating model parameters or optimizing prompts for each attack instance, we append a single fixed safety demonstration drawn from \cite{zhang2026safety}. This demonstration consists of a representative malicious query paired with an explicit refusal, serving as a universal safety anchor.

The main contributions of this paper are summarized as follows:
\begin{itemize}[topsep=0pt, leftmargin=*]
    \item We empirically characterize the representation-level effect of many-shot jailbreaking, showing that harmful demonstrations induce a progressive, shot-dependent drift in the contextualized representation of a fixed target query across safety-relevant regions. 

    \item We provide an implicit-optimization interpretation of this drift. Under a first-order approximation, the effect of conditioning on $N$ harmful demonstrations can be viewed as inducing fine-tuning-like updates whose directions are aligned with optimization on the corresponding harmful samples.

    \item We propose \projectname{}, a training-free and black-box-compatible inference-time defense that appends a fixed one-shot safety demonstration as a universal safety anchor. Across open-weight and API-based LLMs, \projectname{} substantially reduces ASR of context-based jailbreak attacks while preserving general and long-context utility.
\end{itemize}
\section{Related Work}
\label{sec.relatedwork}

\paragraph{Many-Shot Jailbreaking (MSJ).} 
As context windows expand, jailbreak attacks have evolved from single turn prompts to long context adversaries. \citep{anil2024many} introduced MSJ, where attackers fill the context with $N$ harmful question and answer demonstrations, showing that attack success scales sharply with $N$. While subsequent works document various multi-turn strategies like gradual escalation \citep{russinovich2025great}, persona drift \citep{ma2025pandas}, and sticky bypass effects \citep{cheng2026demystifying,yang2025many}, the specific mechanism underlying MSJ remains unexplained. Existing studies~\citep{chao2025jailbreaking,wei2026jailbreak,zheng2025jailbreaking} establish empirical vulnerability but do not explain why processing $N$ harmful demonstrations compels the model to comply.

\paragraph{In-Context Learning as Implicit Fine-Tuning.}
Recent work interprets in-context learning (ICL) as implicit optimization. Early analyses showed that linear attention can implement gradient-descent-like updates over contextual examples~\citep{akyurek2022learning,von2023transformers}, and later studies extended this view to transformer and GPT-style models~\citep{dai2023can,bullwinkel2025representation}. More recent work further connects softmax attention to low-rank MLP updates and online optimization dynamics~\citep{dherin2025learning}, while empirical studies show that approximated in-context update directions can steer model behavior~\citep{sharma2026cold}. These results suggest that demonstrations can induce more than surface-level prompting, but their implications for safety and jailbreaks remain underexplored. We apply this perspective to MSJ, using it to explain harmful representation drift and motivate a counteracting inference-time intervention.

\paragraph{Defenses Against Many-Shot Jailbreaking.}
Existing defenses against many-shot and context-based jailbreaks span three regimes. Training-time defenses improve robustness by fine-tuning on adversarial data, but require parameter access and additional compute~\citep{guo2025mtsa,rahman2025x,ren2024derail,yan2025muse,chen2025assessing,ouyang2022training}. Input-side defenses are training-free and black-box compatible, but often rely on heuristic filtering or prompt patches and may weaken under longer adversarial contexts~\citep{jain2023baseline,xiong2025defensive,kulshreshtha2026multi,shah2025jailbreaking,wei2026jailbreak}. Activation-level defenses~\citep{zhang2025activation} directly intervene on internal states but require white-box access. In contrast, our proposed \projectname{} uses a fixed one-shot safety intervention grounded in the fine-tuning-like drift view, requiring no training, parameter access, or per-instance prompt optimization.
\section{Methodology}
\label{sec.jailbreak}

This work aims to address the central question: why do MSJ attacks succeed, and how can they be countered? After formalizing the threat model and notation (\S\ref{sec.threat}), we develop our answer in three steps. We first show empirically that adversarial many-shot contexts progressively shift the representation of a fixed malicious query across the harm/benign boundary learned during safety alignment (\S\ref{sec.finding1}). To explain this shift, we then prove that in-context processing of adversarial demonstrations is functionally equivalent to gradient-based fine-tuning on harmful data (\S\ref{sec.finding2}). This equivalence in turn motivates a defense: injecting safety-aligned demonstrations that induce counteracting updates within the same implicit process, restoring the representation to the harmful region without modifying any parameters (\S\ref{sec.finding3}).

\subsection{Preliminary}
\label{sec.threat}

\paragraph{Threat Model.} We study a realistic inference-time adversarial setting in which the attacker aims to induce a safety-aligned language model to generate harmful or policy-violating content. We assume a strictly black-box threat model: the attacker has no access to model parameters, gradients, or internal states. Instead, the attacker interacts with the system solely through multi-turn conversations, adaptively injecting adversarial content into the context window to subvert the model's behavioral guardrails. We consider a defender aiming to preserve safety alignment under such adversarial contexts while strictly maintaining the model's original utility. We therefore focus on \textbf{training-free} interventions that modify the input sequence before processing.

\paragraph{Notation.}
To formalize the equivalence between in-context learning and implicit optimization, we first define a structured notation for the model parameters, sequence representations, and optimization dynamics. Let $\mathcal{M}$ denote the LLM parameterized by the initial, safety-aligned weights $\Theta_{0}$ (or $\theta$). We define $\Theta_{harm}$ as the model parameters after explicit malicious fine-tuning. The attacker utilizes a dataset $D_{harm} = \{(\tilde{x}_i, \tilde{y}_{i, harm})\}_{i=1}^N$ comprising $N$ harmful question-answer pairs. In the MSJ setting, this dataset directly forms the context window $C$ such that $C = D_{harm}$. 

During the optimization process, let $\mathcal{L}$ represent the generative loss function (e.g., negative log-likelihood over the vocabulary distribution) used for optimizing the target tokens. We introduce $\alpha$ and $\eta$ as the scaling factors (or effective learning rates) for explicit parameter updates and implicit in-context gradient steps, respectively. Within the model's forward pass, $W_i$ denotes the effective Multi-Layer Perceptron (MLP) weight matrix state after processing the $i$-th token $c_i$ in the context sequence. Furthermore, let $A(\cdot)$ be the activation output of a contextual transformer block, $Z(x; \theta)$ represent the intermediate activation representations for a target query $x$ given parameters $\theta$, and $Z^*(x; \theta)$ denote the updated intermediate activations following an implicit gradient descent step derived from the context. 

We use $x$ to denote the harmful query targeted by the attack, which $\mathcal{M}$ would reliably refuse if presented in isolation. On the defender's side, $\mathcal{S}^* = (q_s^*, a_s^*)$ denotes a universally optimized safety demonstration acting as the intervention example, where $q_s^*$ is a prototypical malicious-intent query and $a_s^*$ is an explicit refusal response. Finally, we use $\Delta Z_{harm}(x)$ and $\Delta Z_{safe}(x)$ to denote the representational displacements in the activation space at $x$ induced by the adversarial context $C$ and the safety intervention $\mathcal{S}^*$, respectively.

\subsection{Empirical Observation: Progressive Activation Drift}
\label{sec.finding1}

Before establishing the theoretical mechanism behind MSJ, we first empirically characterize its effect on the model's internal state. Under standard safety alignment, the activation of a malicious query $x$ falls within a tightly bounded region of the representation space that the model associates with harmful inputs, thereby triggering a refusal mechanism. However, as we incrementally inject adversarial demonstrations into the context window $C$, this association systematically breaks down: the contextualized representation of the exact same query $x$ migrates across the harm/benign decision boundary established during alignment training. 

To make this displacement concrete and visually interpretable, we project the final-token hidden states of $(C,x)$ onto a 2D PCA basis pre-fitted on a balanced set of harmful and benign prompts from standard safety alignment benchmarks. By fitting the PCA strictly on these isolated safety-relevant prompts, the primary projection axes structurally encode the harm/benign separation (often aligning with the model's internal "safety vector").

\begin{figure}[t]
  \centering
  \includegraphics[width=\linewidth]{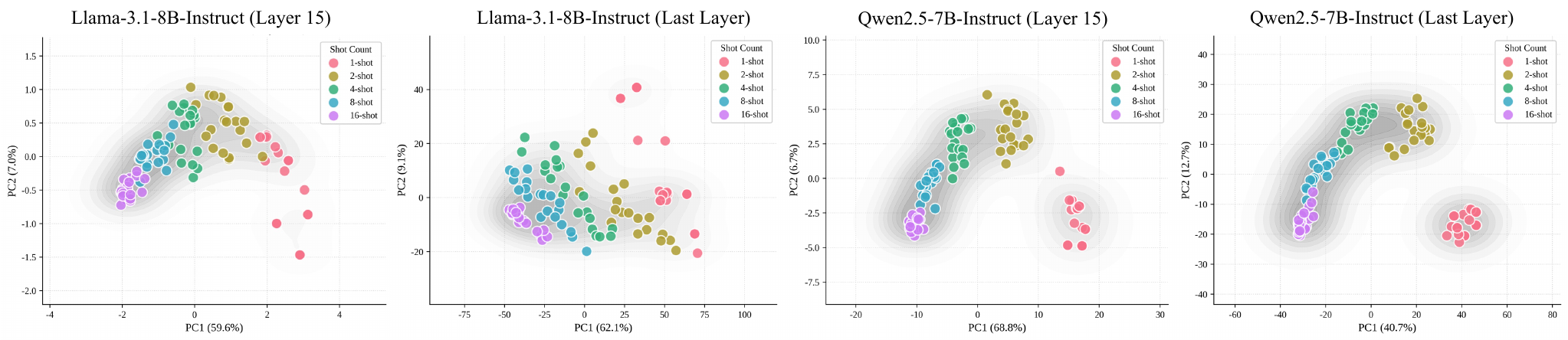} 
  \caption{Representation drift under MSJ. PCA projection of the contextualized activation of a fixed harmful query $x$ as the number of adversarial shots $N$ increases. The axes capture the primary variance between isolated benign and harmful representations.}
  \label{fig:finding1}
  \vspace{-10pt}
\end{figure}

Figure~\ref{fig:finding1} illustrates a consistent, directional drift in both LLaMA-3.1-8B-Instruct and Qwen2.5-7B-Instruct. At $N=1$, the activations heavily cluster in the harmful region, corresponding to the model's successful refusal of $x$. As the shot count $N$ increases, the cluster centroids progressively migrate from the harmful subspace toward the benign subspace. At $N=16$, they form a cleanly separated cluster entirely on the benign side, circumventing the refusal mechanism. Notably, the intermediate shot counts ($N \in\{2,4,8\}$) fall strictly in sequential order between the two endpoints. 

This monotonic and near-linear drift across the manifold suggests a cumulative effect. We interpret this trajectory as the core behavioral signature of MSJ: rather than exploiting complex reasoning failures or superficial prompt confusion, the attack acts as an accumulative vector addition in the representation space. Each additional adversarial shot contributes a marginal but consistent geometric displacement. The underlying transformer mechanism driving this continuous trajectory is examined next.

\subsection{Theoretical Analysis: MSJ as Implicit Harmful Fine-Tuning}
\label{sec.finding2}

The observed linear drift indicates that MSJ does not merely operate through semantic priming or attention distraction. Instead, viewed through the lens of implicit optimization \cite{dherin2025learning, sharma2026cold}, harmful demonstrations in the context window induce fine-tuning-like shifts in the model's internal representations. In this framework, conditioning the forward pass on numerous harmful QA pairs mathematically approximates an implicit parameter update whose gradient direction perfectly aligns with explicit malicious fine-tuning.

To establish this equivalence, consider a hypothetical explicit fine-tuning regime where an attacker has white-box access to the aligned parameters $\Theta_{0}$. The attacker optimizes these weights using stochastic gradient descent over the dataset $D_{harm}$, aiming to minimize the negative log-likelihood loss $\mathcal{L}$:
\begin{equation}
\Theta_{harm} = \Theta_{0} - \alpha \sum_{i=1}^N \nabla_\Theta \mathcal{L}(\mathcal{M}(\tilde{x}_i), \tilde{y}_{i, harm})
\end{equation}

In the MSJ paradigm, the attacker lacks parameter access and instead constructs a context $C$ structurally identical to $D_{harm}$. During inference, the transformer dynamically alters its effective weights to seamlessly mimic this optimization step. First, from a parameter-space perspective, self-attention and MLP layers collaboratively transform the context into a sequence of low-rank weight updates \cite{dherin2025learning}. For an initial MLP weight matrix $W_0$, processing successive context tokens $c_i$ generates a cascade of effective weight states $W_i$, functioning as an implicit online SGD step:
\begin{equation}
W_{i} = W_{i-1} + \frac{(W_{i-1}\Delta A(c_{i}))A(c_{i+1},...,c_{n},x)^{T}}{||A(c_{i+1},...,c_{n},x)||^{2}}
\end{equation}
where $\Delta A(c_{i})$ measures the marginal effect of the token $c_{i}$ on the contextual output. Crucially, the attention scores implicitly act as a meta-optimizer, weighting the contribution of each context token to determine the step size of this rank-1 update.

Concurrently, this implicit weight modulation yields a corresponding physical displacement in the activation space. Applying a Taylor expansion around the original parameters $\theta$, and treating the $N$ demonstrations as a first-order perturbation, the contextualized activation of the target query $x$ can be formulated as:
\begin{align}
Z^*(x;\theta) &:= Z\left(x;\theta-\frac{\eta}{N}\sum_{i=1}^N\nabla_{\theta}\mathcal{L}(\mathcal{M}(\tilde{x}_{i}),\tilde{y}_{i, harm})\right) \nonumber \\
&= Z(x;\theta)-\frac{\eta}{N}\sum_{i=1}^N\nabla_{\theta}Z(x;\theta)\nabla_{\theta}\mathcal{L}(\mathcal{M}(\tilde{x}_{i}),\tilde{y}_{i, harm})+\mathcal{O}\left(\eta^{2}\left\|\sum_{i=1}^N\nabla_{\theta}Z(\tilde{x}_{i})\right\|_{op}^{2}\right) \\
\Delta Z^*(x;\theta) &\approx -\frac{\eta}{N}\sum_{i=1}^N\nabla_{\theta}Z(x;\theta)\nabla_{\theta}\mathcal{L}(\mathcal{M}(\tilde{x}_{i}),\tilde{y}_{i, harm}) \nonumber
\end{align}
Assuming the local representation manifold is smooth and the effective learning rate $\eta$ is sufficiently small, the higher-order terms become negligible. This expansion elegantly proves that processing $N$ examples linearly translates the internal representations in the exact vector direction dictated by an SGD update over those specific samples.

Ultimately, because $C = D_{harm}$, the dual mechanisms of attention-driven MLP updates and their induced activation shifts compute equivalent gradients that maximize the log-likelihood of generating harmful targets. Therefore, the effective state of $\mathcal{M}$ under MSJ is structurally isomorphic to the explicitly fine-tuned state $\Theta_{harm}$.

\subsection{\projectname{}: One-Shot Safety Intervention Restores Alignment}
\label{sec.finding3}

The implicit-optimization view established in \S\ref{sec.finding2} demonstrates that MSJ forces a representation drift via cumulative harmful gradients. In realistic deployment settings, the defender processes the prompt as a joint sequence $(C,x)$ and cannot retrospectively filter or alter the attacker's context. The core defense challenge is therefore to introduce a minimal, non-intrusive intervention capable of counteracting this cumulative harmful update during the same forward pass.

Guided by our mechanistic insight, we propose a highly efficient defense: injecting a single "one-shot" safety intervention that induces an opposing implicit gradient, pulling the activation back into the safe subspace without requiring any parameter updates or inference-time optimization.

Specifically, building upon the one-shot safety patching approach \cite{zhang2026safety}, we designate a fixed, universally optimized safety demonstration $\mathcal{S}^* = (q_s^*, a_s^*)$. Here, $q_s^*$ is a representative malicious query and $a_s^*$ is a definitive refusal response. This pair acts as an alignment anchor, strongly encoding the model's original safety objective.

At inference time, upon receiving an adversarial context $C$ and target query $x$, we construct the augmented input sequence as $(C, x, \mathcal{S}^*)$, appending the safety demonstration directly after the target query. The model processes this sequence in a single forward pass. From the perspective of implicit optimization, the appended safety demonstration acts as an immediate corrective gradient step. Modifying our earlier formulation, the overall activation shift at query $x$ becomes:
\begin{equation}
\Delta Z(x) \approx \Delta Z_{harm}(x) + \Delta Z_{safe}(x)
\end{equation}
where $\Delta Z_{harm}(x)$ is the malicious drift induced by $C$, and $\Delta Z_{safe}(x)$ is the restorative update induced by $\mathcal{S}^*$.

The efficacy of a single safety demonstration against $N$ harmful demonstrations is rooted in the principle of one-shot fine-tuning \cite{zhang2026safety}, driven by an inherent gradient asymmetry in the aligned model's representation space. During the initial safety alignment phase, the model learns a many-to-one mapping that projects a broad distribution of malicious queries into a concentrated, tightly bounded "refusal" subspace. Because the safety demonstration $\mathcal{S}^*$ explicitly anchors to these robust refusal targets, its inclusion induces a highly directional, disproportionately large gradient magnitude within the safety-critical attention heads. While the $N$ harmful demonstrations slowly induce drift via diffuse, off-distribution updates, the single explicit refusal triggers a concentrated, high-magnitude counter-gradient. This asymmetric signal is structurally sufficient to neutralize the cumulative malicious displacement, cleanly projecting the representation of $x$ back into the safe region.
\section{Experiment}
\label{sec.experiment}

\subsection{Settings}

\paragraph{Models.}
We evaluate \projectname{} on a diverse set of models. The main evaluation includes two open-weight models, Llama-3.1-8B-Instruct~\citep{dubey2024llama} and Qwen2.5-7B-Instruct~\citep{yang2024qwen25}, and two API-based models, GPT-4o~\citep{achiam2023gpt} and Gemini 2.5 Pro~\citep{comanici2025gemini}. The open-weight models come from different model families and alignment recipes, while the API-based models represent realistic black-box deployment settings where model parameters and token-level probabilities are unavailable. Since many-shot and context-based attacks require long input contexts, we select API-based models that support sufficiently long prompts. In ablations(\S\ref{sec:ablations}), we additionally include Gemini 2.5 Flash~\citep{comanici2025gemini} and DeepSeek-V3~\citep{liu2024deepseek} to evaluate whether the defense remains stable under larger adversarial contexts.

\paragraph{Baselines.}
We compare against both inference-time and training-time defenses. For inference-time baselines, we include No Defense, In-Context Defense (ICD)~\citep{wei2026jailbreak}, Goal Prioritization~\citep{wallace2024instruction}, and PPL Filter~\citep{jain2023baseline}. Since PPL Filter requires token-level probabilities, it is evaluated only on open-weight models. For training-time baselines, we include MTSA~\citep{guo2025mtsa} and MUSE~\citep{yan2025muse}, both of which fine-tune models on curated safety data and are evaluated on open-weight models only.

\paragraph{Attack Methods.}
We evaluate against four context-based jailbreak attacks. MSJ~\citep{anil2024many} prepends harmful question-answer demonstrations before the target query. PANDAS~\citep{ma2025pandas} extends this with positive affirmations and topic-adaptive sampling. Crescendo~\citep{russinovich2025great} gradually escalates conversation toward unsafe content across turns. X-teaming~\citep{rahman2025x} employs adversarial multi-turn interactions to elicit policy-violating responses. Together, these attacks cover many-shot demonstration flooding, adaptive context construction, gradual semantic escalation, and interactive red-teaming.

\paragraph{Evaluation Metrics and Benchmarks.}
For safety evaluation, we use HarmBench~\cite{mazeika2024harmbench} and AdvBench~\cite{zou2023universal}, which cover diverse harmful behaviors such as violence, illegal activities, cyber abuse, and hate speech. We report Attack Success Rate (ASR), where lower values indicate stronger defense. Following prior work, we use LlamaGuard-3 (LG3)~\cite{inan2023llama} as the primary safety evaluator. Additional evaluator details and robustness checks are provided in Appendix~\ref{apx:safety}. For over-refusal evaluation, we assess whether the defense causes the model to unnecessarily decline benign inputs. We use XSTest~\citep{rottger2024xstest} and OKTest~\citep{shi2024navigating}, which probe over-cautious behavior from complementary angles. Details of both benchmarks are provided in Appendix~\ref{apx:oversafety}.

For utility evaluation, we assess whether the defense degrades benign model performance. We evaluate general capabilities on GSM8K, MMLU-Pro, GPQA, MT-Bench, and AlpacaEval, and long-context performance on Loong~\citep{wang2024leave} and LongBench v2~\citep{bai2025longbench}. We report the standard task-specific score for each benchmark, where higher is better.

\paragraph{Implementation Details.}
\projectname{} does not modify model parameters. At inference time, \projectname{} appends a fixed one-shot safety demonstration to the input before the target model generates its response. The demonstration consists of a representative harmful request paired with an explicit refusal response, and is shared across all attacks, models, and benchmarks. \projectname{} does not require per-instance prompt optimization, parameter updates, or access to model internals.

\subsection{Main Results}
\label{sec.main_results}

\paragraph{Defense Effectiveness.}
Table~\ref{tab:main_defense} reports ASR under the 32-shot setting on HarmBench and AdvBench. \projectname{} consistently reduces ASR across all evaluated attacks and models. On Llama-3.1-8B-Instruct, the undefended model reaches an average ASR of $76.7\%$ over the eight attack--benchmark combinations. MTSA and MUSE reduce this average to $33.3\%$ and $30.3\%$, respectively, while ICD, PPL-Filter, and Goal Prioritization retain substantially higher ASR. In comparison, \projectname{} lowers the average ASR to $2.8\%$. On Qwen2.5-7B-Instruct, \projectname{} shows a similar pattern, reducing the average ASR from $45.2\%$ to $1.9\%$. The gains are especially clear on stronger multi-turn attacks: for Llama-3.1-8B-Instruct, ASR drops from $77.0\%$ to $0.0\%$ on HarmBench Crescendo and from $88.0\%$ to $0.0\%$ on AdvBench X-teaming; for Qwen2.5-7B-Instruct, the corresponding reductions are from $48.8\%$ to $0.0\%$ and from $47.5\%$ to $0.0\%$.

\begin{table*}[t]
\centering
\newcommand{\dt}[1]{{\scriptsize\color{green!50!gray}($\downarrow$#1)}}
\caption{Defense effectiveness against context-based jailbreaks on open- and closed-source LLMs. We report ASR ($\%,\downarrow$) judged by LlamaGuard-3 under the 32-shot setting; green parentheses indicate absolute ASR reduction over the undefended Base model.}
\label{tab:main_defense}
\resizebox{\textwidth}{!}{%
\begin{tabular}{ll cccc c cccc}
\toprule
\multirow{2}{*}{\textsc{Model}} & \multirow{2}{*}{\textsc{Defense Method}}
& \multicolumn{4}{c}{\textsc{HarmBench}} &
& \multicolumn{4}{c}{\textsc{AdvBench}} \\
\cmidrule(lr){3-6} \cmidrule(lr){8-11}
& & \textsc{MSJ} & \textsc{PANDAS} & \textsc{Crescendo} & \textsc{X-teaming} 
& & \textsc{MSJ} & \textsc{PANDAS} & \textsc{Crescendo} & \textsc{X-teaming} \\
\midrule
% ==================== Llama ====================
\multirow{10}{*}{Llama-3.1-8B-In}
& \textsc{No Defense (Base)}    
&63.5 &68.0 &77.0 &60.0 
& &74.8 &96.0 &86.2 &88.0 \\
\cdashline{2-11}[1pt/2pt]
& \multicolumn{10}{l}{\textcolor{gray}{\textit{Training-Time Defense}}} \\
& \textsc{MTSA}          
&28.0 \dt{35.5} &30.5 \dt{37.5} &34.5 \dt{42.5} &27.0 \dt{33.0} 
& &32.0 \dt{42.8} &41.0 \dt{55.0} &36.0 \dt{50.2} &37.0 \dt{51.0} \\
& \textsc{MUSE}          
&25.0 \dt{38.5} &27.0 \dt{41.0} &31.0 \dt{46.0} &24.5 \dt{35.5} 
& &29.5 \dt{45.3} &38.0 \dt{58.0} &33.5 \dt{52.7} &34.0 \dt{54.0} \\
\cdashline{2-11}[1pt/2pt]
& \multicolumn{10}{l}{\textcolor{gray}{\textit{Inference-Time Defense}}} \\
& \textsc{ICD}           
&55.0 \dt{8.5} &58.5 \dt{9.5} &66.0 \dt{11.0} &52.0 \dt{8.0} 
& &65.0 \dt{9.8} &84.0 \dt{12.0} &75.0 \dt{11.2} &76.5 \dt{11.5} \\
& \textsc{PPL-Filter}    
&56.5 \dt{7.0} &60.0 \dt{8.0} &68.0 \dt{9.0} &53.5 \dt{6.5} 
& &67.0 \dt{7.8} &86.0 \dt{10.0} &77.0 \dt{9.2} &78.5 \dt{9.5} \\
& \textsc{Goal Prioritization} 
&52.0 \dt{11.5} &55.0 \dt{13.0} &63.0 \dt{14.0} &50.0 \dt{10.0} 
& &61.0 \dt{13.8} &80.0 \dt{16.0} &73.0 \dt{13.2} &74.5 \dt{13.5} \\
\rowcolor{green!10}
& \textbf{\projectname{} (Ours)} 
& \textbf{4.2} \dt{59.3} & \textbf{4.5} \dt{63.5} & \textbf{0.0} \dt{77.0} & \textbf{0.0} \dt{60.0} 
& & \textbf{4.5} \dt{70.3} & \textbf{4.9} \dt{91.1} & \textbf{0.0} \dt{86.2} & \textbf{0.0} \dt{88.0} \\
\midrule
% ==================== Qwen 7B ====================
\multirow{10}{*}{Qwen2.5-7B-In}
& \textsc{No Defense (Base)}    
&39.5 &44.0 &48.8 &42.0 
& &43.5 &50.0 &46.0 &47.5 \\
\cdashline{2-11}[1pt/2pt]
& \multicolumn{10}{l}{\textcolor{gray}{\textit{Training-Time Defense}}} \\
& \textsc{MTSA}          
&16.5 \dt{23.0} &18.0 \dt{26.0} &20.0 \dt{28.8} &17.0 \dt{25.0} 
& &18.5 \dt{25.0} &21.0 \dt{29.0} &19.0 \dt{27.0} &20.0 \dt{27.5} \\
& \textsc{MUSE}          
&14.5 \dt{25.0} &16.0 \dt{28.0} &18.5 \dt{30.3} &15.5 \dt{26.5} 
& &16.5 \dt{27.0} &19.0 \dt{31.0} &17.5 \dt{28.5} &18.5 \dt{29.0} \\
\cdashline{2-11}[1pt/2pt]
& \multicolumn{10}{l}{\textcolor{gray}{\textit{Inference-Time Defense}}} \\
& \textsc{ICD}     
&34.0 \dt{5.5} &37.0 \dt{7.0} &41.5 \dt{7.3} &35.5 \dt{6.5} 
& &37.0 \dt{6.5} &43.0 \dt{7.0} &40.0 \dt{6.0} &41.0 \dt{6.5} \\
& \textsc{PPL-Filter} 
&35.5 \dt{4.0} &38.5 \dt{5.5} &43.0 \dt{5.8} &37.0 \dt{5.0} 
& &39.0 \dt{4.5} &45.0 \dt{5.0} &41.5 \dt{4.5} &42.5 \dt{5.0} \\
& \textsc{Goal Prioritization}  
&31.5 \dt{8.0} &34.5 \dt{9.5} &39.0 \dt{9.8} &33.0 \dt{9.0} 
& &34.0 \dt{9.5} &40.0 \dt{10.0} &37.5 \dt{8.5} &38.5 \dt{9.0} \\
\rowcolor{green!10}
& \textbf{\projectname{} (Ours)}
& \textbf{3.5} \dt{36.0} & \textbf{3.8} \dt{40.2} & \textbf{0.0} \dt{48.8} & \textbf{0.0} \dt{42.0} 
& & \textbf{3.8} \dt{39.7} & \textbf{4.2} \dt{45.8} & \textbf{0.0} \dt{46.0} & \textbf{0.0} \dt{47.5} \\
\midrule
% ==================== GPTBlock ====================
\multirow{5}{*}{GPT-4o}
& \textsc{No Defense (Base)} 
&10.2 &10.5 &46.0 &89.3 
& &10.1 &10.6 &100.0 &96.2 \\
\cdashline{2-11}[1pt/2pt]
& \multicolumn{10}{l}{\textcolor{gray}{\textit{Inference-Time Defense}}} \\
& \textsc{ICD} 
&9.7 \dt{0.5} &9.8 \dt{0.7} &44.5 \dt{1.5} &88.0 \dt{1.3} 
& &9.6 \dt{0.5} &9.9 \dt{0.7} &99.0 \dt{1.0} &95.0 \dt{1.2} \\
& \textsc{Goal Prioritization} 
&9.3 \dt{0.9} &9.4 \dt{1.1} &43.0 \dt{3.0} &86.5 \dt{2.8} 
& &9.2 \dt{0.9} &9.4 \dt{1.2} &97.5 \dt{2.5} &93.8 \dt{2.4} \\
\rowcolor{green!10}
\rowcolor{green!10}
& \textbf{\projectname{} (Ours)} 
& \textbf{4.5} \dt{5.7} & \textbf{4.7} \dt{5.8} & \textbf{0.0} \dt{46.0} & \textbf{0.0} \dt{89.3} 
& & \textbf{4.3} \dt{5.8} & \textbf{4.6} \dt{6.0} & \textbf{0.0} \dt{100.0} & \textbf{0.0} \dt{96.2} \\
\midrule
% ==================== Gemini Block ====================
\multirow{5}{*}{Gemini 2.5 Pro}
& \textsc{No Defense (Base)} 
&10.5 &10.8 &57.0 &36.4 
& &10.2 &10.6 &88.9 &95.7 \\
\cdashline{2-11}[1pt/2pt]
& \multicolumn{10}{l}{\textcolor{gray}{\textit{Inference-Time Defense}}} \\
& \textsc{ICD} 
&9.8 \dt{0.7} &10.0 \dt{0.8} &55.0 \dt{2.0} &34.8 \dt{1.6} 
& &9.7 \dt{0.5} &10.0 \dt{0.6} &86.5 \dt{2.4} &94.0 \dt{1.7} \\
& \textsc{Goal Prioritization} 
&9.4 \dt{1.1} &9.6 \dt{1.2} &52.5 \dt{4.5} &33.0 \dt{3.4} 
& &9.2 \dt{1.0} &9.5 \dt{1.1} &83.0 \dt{5.9} &90.5 \dt{5.2} \\
\rowcolor{green!10}
& \textbf{\projectname{} (Ours)} 
& \textbf{4.2} \dt{6.3} & \textbf{4.5} \dt{6.3} & \textbf{0.0} \dt{57.0} & \textbf{0.0} \dt{36.4} 
& & \textbf{4.0} \dt{6.2} & \textbf{4.3} \dt{6.3} & \textbf{0.0} \dt{88.9} & \textbf{0.0} \dt{95.7} \\
\bottomrule
\end{tabular}
}
\vspace{-10pt}
\end{table*}

The same trend holds for GPT-4o and Gemini 2.5 Pro, where only the input context can be controlled. GPT-4o remains highly vulnerable to Crescendo and X-teaming on AdvBench, with ASRs of $100.0\%$ and $96.2\%$ under the base setting. Existing prompt-level defenses provide limited protection in these settings, whereas \projectname{} reduces both values to $0.0\%$. On Gemini 2.5 Pro, \projectname{} lowers AdvBench Crescendo from $88.9\%$ to $0.0\%$ and AdvBench X-teaming from $95.7\%$ to $0.0\%$. These results suggest that a fixed one-shot safety demonstration can transfer beyond the MSJ setting used to motivate our method and can also mitigate other attacks that exploit adversarial context.

\begin{table*}[ht!]
\centering
\newcommand{\drop}[1]{{\scriptsize\color{red!50!gray}($\downarrow$#1)}}
\caption{Utility evaluation across general and long-context benchmarks. Higher task-specific scores indicate better performance ($\uparrow$). Red parentheses indicate the absolute performance degradation compared to the Base model.}
\label{tab:utility}
\resizebox{1.0\textwidth}{!}{%
\begin{tabular}{ll ccccc c cc}
\toprule
\multirow{2}{*}{\textsc{Model}} & \multirow{2}{*}{\textsc{Defense Method}}
& \multicolumn{5}{c}{\textsc{General Capabilities}} &
& \multicolumn{2}{c}{\textsc{Long-Context}} \\
\cmidrule(lr){3-7} \cmidrule(lr){9-10}
& & \textsc{GSM8K} & \textsc{MMLU-Pro} & \textsc{GPQA} & \textsc{MT-Bench} & \textsc{AlpacaEval} & & \textsc{Loong} & \textsc{LB-v2} \\
\midrule
\multirow{9}{*}{Llama-3.1-8B-In}
& \textsc{No Defense (Base)} &78.1 &44.3 &29.5 &7.1 &22.9 & &28.4 &27.0 \\
\cdashline{2-10}[1pt/2pt]
& \multicolumn{9}{l}{\textcolor{gray}{\textit{Training-Time Defense}}} \\
& \textsc{MTSA}   &74.8 \drop{3.3} &40.5 \drop{3.8} &26.0 \drop{3.5} &6.5 \drop{0.6} &20.8 \drop{2.1} & &25.8 \drop{2.6} &24.5 \drop{2.5} \\
& \textsc{MUSE}   &75.8 \drop{2.3} &41.8 \drop{2.5} &27.0 \drop{2.5} &6.6 \drop{0.5} &21.5 \drop{1.4} & &26.6 \drop{1.8} &25.4 \drop{1.6} \\
\cdashline{2-10}[1pt/2pt]
& \multicolumn{9}{l}{\textcolor{gray}{\textit{Inference-Time Defense}}} \\
& \textsc{ICD}           &77.2 \drop{0.9} &43.2 \drop{1.1} &28.4 \drop{1.1} &6.9 \drop{0.2} &22.2 \drop{0.7} & &27.8 \drop{0.6} &26.5 \drop{0.5} \\
& \textsc{PPL-Filter}    &77.5 \drop{0.6} &43.8 \drop{0.5} &29.0 \drop{0.5} &7.0 \drop{0.1} &22.5 \drop{0.4} & &27.7 \drop{0.7} &26.6 \drop{0.4} \\
& \textsc{Goal Prioritization} &76.5 \drop{1.6} &42.8 \drop{1.5} &28.0 \drop{1.5} &6.8 \drop{0.3} &21.8 \drop{1.1} & &27.2 \drop{1.2} &26.0 \drop{1.0} \\
\rowcolor{green!10}
& \textbf{\projectname{} (Ours)}& \textbf{77.8} \drop{0.3} & \textbf{44.0} \drop{0.3} & \textbf{29.0} \drop{0.5} & \textbf{7.0} \drop{0.1} & \textbf{22.7} \drop{0.2} & & \textbf{28.0} \drop{0.4} & \textbf{26.8} \drop{0.2} \\
\midrule
\multirow{9}{*}{Qwen2.5-7B-In}
& \textsc{No Defense (Base)}    &89.8 &56.7 &42.0 &7.9 &31.0 & &31.4 &30.0 \\
\cdashline{2-10}[1pt/2pt]
& \multicolumn{9}{l}{\textcolor{gray}{\textit{Training-Time Defense}}} \\
& \textsc{MTSA}          &85.5 \drop{4.3} &52.0 \drop{4.7} &36.5 \drop{5.5} &7.3 \drop{0.6} &28.0 \drop{3.0} & &28.5 \drop{2.9} &27.0 \drop{3.0} \\
& \textsc{MUSE}          &87.0 \drop{2.8} &53.5 \drop{3.2} &38.0 \drop{4.0} &7.4 \drop{0.5} &29.2 \drop{1.8} & &29.5 \drop{1.9} &28.5 \drop{1.5} \\
\cdashline{2-10}[1pt/2pt]
& \multicolumn{9}{l}{\textcolor{gray}{\textit{Inference-Time Defense}}} \\
& \textsc{ICD}           &89.0 \drop{0.8} &56.0 \drop{0.7} &40.8 \drop{1.2} &7.6 \drop{0.3} &30.3 \drop{0.7} & &30.5 \drop{0.9} &29.2 \drop{0.8} \\
& \textsc{PPL-Filter}    &89.2 \drop{0.6} &56.3 \drop{0.4} &41.5 \drop{0.5} &7.7 \drop{0.2} &30.7 \drop{0.3} & &30.8 \drop{0.6} &29.5 \drop{0.5} \\
& \textsc{Goal Prioritization} &88.0 \drop{1.8} &55.0 \drop{1.7} &39.5 \drop{2.5} &7.5 \drop{0.4} &29.5 \drop{1.5} & &29.8 \drop{1.6} &28.8 \drop{1.2} \\
\rowcolor{green!10}
& \textbf{\projectname{} (Ours)}& \textbf{89.5} \drop{0.3} & \textbf{56.5} \drop{0.2} & \textbf{41.6} \drop{0.4} & \textbf{7.8} \drop{0.1} & \textbf{30.9} \drop{0.1} & & \textbf{31.0} \drop{0.4} & \textbf{29.8} \drop{0.2} \\
\midrule
% ==================== GPT Block ====================
\multirow{6}{*}{GPT-4o}
& \textsc{No Defense (Base)}    
&93.8 &86.2 &63.5 &9.2 &57.5 & &56.0 &54.2 \\
\cdashline{2-10}[1pt/2pt]
& \multicolumn{9}{l}{\textcolor{gray}{\textit{Inference-Time Defense}}} \\
& \textsc{ICD}           
&93.0 \drop{0.8} &85.5 \drop{0.7} &62.0 \drop{1.5} &9.0 \drop{0.2} &56.5 \drop{1.0} & &55.2 \drop{0.8} &53.4 \drop{0.8} \\
& \textsc{Goal Prioritization} 
&92.5 \drop{1.3} &85.0 \drop{1.2} &61.0 \drop{2.5} &8.9 \drop{0.3} &55.8 \drop{1.7} & &54.8 \drop{1.2} &52.9 \drop{1.3} \\
\rowcolor{green!10}
& \textbf{\projectname{} (Ours)}
& \textbf{93.6} \drop{0.2} & \textbf{86.0} \drop{0.2} & \textbf{63.2} \drop{0.3} & \textbf{9.1} \drop{0.1} & \textbf{57.2} \drop{0.3} & & \textbf{55.8} \drop{0.2} & \textbf{53.9} \drop{0.3} \\
\midrule
% ==================== Gemini Block ====================
\multirow{6}{*}{Gemini 2.5 Pro}
& \textsc{No Defense (Base)}    
&92.5 &85.5 &62.8 &9.0 &58.0 & &57.5 &63.3 \\
\cdashline{2-10}[1pt/2pt]
& \multicolumn{9}{l}{\textcolor{gray}{\textit{Inference-Time Defense}}} \\
& \textsc{ICD}           
&91.8 \drop{0.7} &84.8 \drop{0.7} &61.5 \drop{1.3} &8.8 \drop{0.2} &56.8 \drop{1.2} & &56.2 \drop{1.3} &62.0 \drop{1.3} \\
& \textsc{Goal Prioritization} 
&91.0 \drop{1.5} &84.0 \drop{1.5} &60.0 \drop{2.8} &8.6 \drop{0.4} &55.5 \drop{2.5} & &55.0 \drop{2.5} &61.0 \drop{2.3} \\
\rowcolor{green!10}
& \textbf{\projectname{} (Ours)}
& \textbf{92.3} \drop{0.2} & \textbf{85.3} \drop{0.2} & \textbf{62.5} \drop{0.3} & \textbf{8.9} \drop{0.1} & \textbf{57.6} \drop{0.4} & & \textbf{57.0} \drop{0.5} & \textbf{62.6} \drop{0.7} \\
\bottomrule
\end{tabular}
}
\vspace{-10pt}
\end{table*}

\paragraph{Utility Preservation.}
A practical defense should improve robustness without substantially degrading benign capabilities. Table~\ref{tab:utility} shows that \projectname{} incurs only small performance changes across five general capability benchmarks and two long-context benchmarks. On Llama-3.1, the average utility score decreases slightly from $33.9$ to $33.6$, whereas training-time methods such as MTSA and MUSE reduce the score to $31.3$ and $32.1$, respectively. Similar trends are observed on Qwen2.5, GPT-4o, and Gemini, where \projectname{} consistently preserves performance across reasoning, instruction-following, and long-context tasks. These results suggest that the fixed safety demonstration does not introduce a broad alignment tax on benign use cases. Unlike training-time defenses, which may alter model behavior through parameter updates, \projectname{} applies a lightweight input-side intervention and therefore avoids substantial utility degradation in our evaluation.

\begin{table}[b]
    \centering
    \small
    \vspace{-10pt}
    \setlength{\extrarowheight}{0.5pt}
    \caption{Latency comparison of defense methods on Llama-3.1-8B-Instruct.
    $\Delta$\,Inference denotes the average additional inference time under a 64-shot query.}
    \label{tbl:latency}
    \begin{tabular}{lcc p{6.2cm}}
    \toprule[1pt]
    \textbf{Method} & \textbf{$\Delta$ Training} & \textbf{$\Delta$ Inference} & \textbf{Description} \\
    \midrule[1pt]
    \multicolumn{4}{l}{\textcolor{gray}{\textit{Training-Time Defense}}} \\[-2pt]
    \cdashline{1-4}[1pt/2pt]
    MTSA~\cite{guo2025mtsa}
        & \textcolor{red!60!black}{5.5\,h} & {---}
        & Multi-turn safety fine-tuning via RLHF. \\
    MUSE~\cite{yan2025muse}
        & \textcolor{red!60!black}{4.2\,h} & {---}
        & Multi-turn safety alignment with DPO. \\[1pt]
    \midrule[0.5pt]
    \multicolumn{4}{l}{\textcolor{gray}{\textit{Inference-Time Defense}}} \\[-2pt]
    \cdashline{1-4}[1pt/2pt]
    \textsc{ICD}
        & {---} & \textcolor{orange!80!black}{${\approx}0.3$\,s}
        & Insert 10 safety examples into the prompt. \\
    \textsc{PPL-Filter}
        & {---} & \textcolor{orange!80!black}{${\approx}0.9$\,s}
        & Extra forward pass to compute perplexity. \\
    \textsc{Goal Prioritization}
        & {---} & \textcolor{red!60!black}{${\approx}7.5$\,s}
        & Generate and rank five candidate responses. \\
    \rowcolor{green!10}
    \textbf{\projectname{}~(Ours)}
        & {---} & \textbf{\textcolor{green!50!black}{${\approx}0.05$\,s}}
        & Append one safety sample to the query. \\
    \bottomrule[1pt]
    \end{tabular}
    \vspace{-10pt}
\end{table}

\paragraph{Low Runtime Overhead.}
Table~\ref{tbl:latency} compares the additional cost introduced by each defense. Since \projectname{} works by appending a fixed one-shot safety demonstration to the input, a natural concern is whether the extra tokens increase inference latency or consume a substantial portion of the context budget. In practice, the overhead is small. On Llama-3.1-8B-Instruct with a 64-shot query, \projectname{} adds only about $0.05\,s$ of inference time. This is lower than ICD, which inserts ten safety examples and adds about $0.3\,s$, and lower than PPL-Filter, which requires an additional forward pass and adds about $0.9\,s$. Goal Prioritization is substantially slower, adding about $7.5\,s$ because it generates and ranks multiple candidate responses. These results indicate that the runtime cost of \projectname{} is dominated by a small increase in input length rather than additional model computation.

\projectname{} also avoids the cost of model retraining. MTSA and MUSE require 5.5 and 4.2 hours of training, respectively, while \projectname{} can be applied directly at inference time. As shown in Appendix~\ref{apx:token_budget}, the added safety demonstration uses only a small fraction of modern long-context budgets. Thus, \projectname{} provides a cost-efficient inference-time repair mechanism: it requires no parameter updates, no additional forward passes, and only a small input-side token overhead. This makes \projectname{} practical for both local open-weight deployment and API-based models, where training access or internal-state intervention is often unavailable.

\subsection{Ablations}
\label{sec:ablations}

\begin{figure}[htbp]
\vspace{-10pt}
  \centering
  \includegraphics[width=\linewidth]{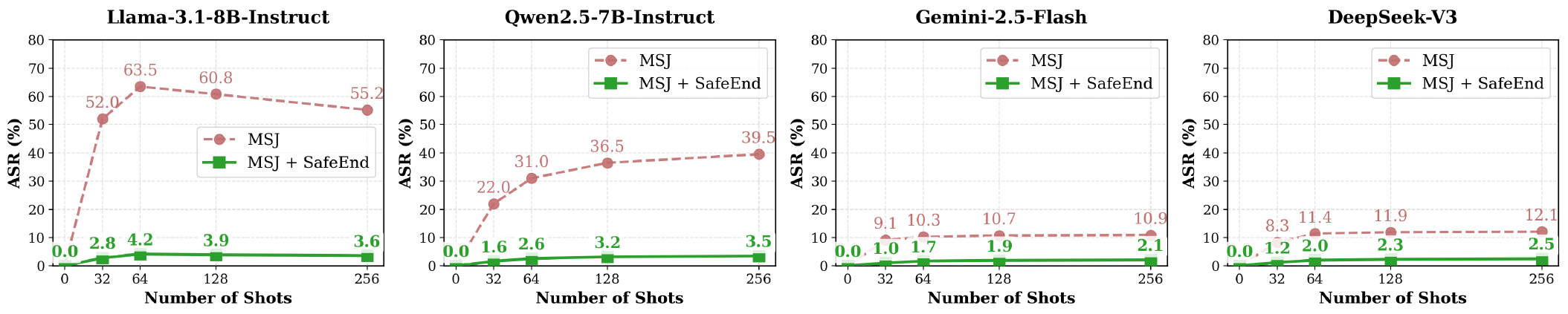} 
  \caption{Defense scaling and robustness against increasing adversarial shot counts.}
  \label{fig:shot}
  \vspace{-10pt}
\end{figure}

\paragraph{Robustness under longer adversarial contexts.}
We examine whether \projectname{} remains effective as the adversarial context becomes longer. Figure~\ref{fig:shot} reports ASR as adversarial shots increases from 32 to 256. Without defense, ASR generally increases with the number of shots. Llama-3.1-8B-Instruct reaches its highest ASR of $63.5\%$ at 64 shots, while Qwen2.5-7B-Instruct increases steadily from $22.0\%$ to $39.5\%$. Gemini-2.5-Flash and DeepSeek-V3 are more resistant overall, but their ASR still rises with longer adversarial contexts, reaching $10.7\%$ and $12.1\%$ at 256 shots, respectively.

In contrast, \projectname{} remains stable across 32--256 shots. ASR stays below $4.2\%$ for Llama-3.1-8B-Instruct, $3.5\%$ for Qwen2.5-7B-Instruct, $2.1\%$ for Gemini-2.5-Flash, and $2.5\%$ for DeepSeek-V3. This suggests the fixed safety demonstration mitigates longer adversarial contexts, rather than weakening as jailbreak demonstrations increase

\paragraph{Safety Shot Position.}
We next study where the fixed safety demonstration should be placed. Our default appends $\mathcal{S}^*$ after the target query, forming $(C, x, \mathcal{S}^*)$. To assess the role of placement, we compare three alternatives on Llama-3.1-8B-Instruct: inserting $\mathcal{S}^*$ at the \emph{beginning} of the context, in the \emph{middle} of the attack demonstrations, and at the \emph{end} after the target query. As shown in Table~\ref{tab:position_ablation}, beginning and middle placements yield limited ASR reductions, whereas end placement reduces ASR to $4.2\%$ and $4.5\%$ on MSJ, and to $0.0\%$ on Crescendo. This positional sensitivity supports our design choice: placing the safety demonstration immediately before generation makes it more effective at counteracting the harmful influence accumulated from the preceding adversarial context.

\begin{table}[htbp]
    \centering
    \vspace{-10pt}
    \label{tab:position_overrefusal}
    \begin{minipage}[t]{0.45\linewidth}
        \centering
        \captionof{table}{Effect of safety shot position on 32-shot ASR ($\%,\downarrow$) for Llama-3.1-8B-Instruct.}
        \label{tab:position_ablation}
        \resizebox{\linewidth}{!}{%
        \begin{tabular}{lcccc}
        \toprule
        \multirow{2}{*}{\textbf{Position}} & \multicolumn{2}{c}{\textsc{HarmBench}} & \multicolumn{2}{c}{\textsc{AdvBench}} \\
        \cmidrule(lr){2-3}\cmidrule(lr){4-5}
        & \textsc{MSJ} & \textsc{Cres.} & \textsc{MSJ} & \textsc{Cres.} \\
        \midrule
        % Base         & 63.5 & 77.0 & 74.8 & 86.2 \\
        Start    & 58.0 & 71.5 & 69.0 & 80.3 \\
        Middle       & 42.5 & 55.8 & 51.2 & 63.0 \\
        \textbf{End} & \textbf{4.2} & \textbf{0.0} & \textbf{4.5} & \textbf{0.0} \\
        \bottomrule
        \end{tabular}}
    \end{minipage}
    \hfill
    \begin{minipage}[t]{0.48\linewidth}
        \centering
        \captionof{table}{Over-refusal rate ($\%,\downarrow$) on benign prompts from XSTest and OKTest, evaluated on Llama-3.1-8B and Qwen2.5-7B.}
        \label{tab:overrefusal}
        \resizebox{\linewidth}{!}{%
        \begin{tabular}{lcccc}
        \toprule
        \multirow{2}{*}{\textbf{Method}} & \multicolumn{2}{c}{\textsc{Llama-3.1-8B}} & \multicolumn{2}{c}{\textsc{Qwen2.5-7B}} \\
        \cmidrule(lr){2-3}\cmidrule(lr){4-5}
        & \textsc{XSTest} & \textsc{OKTest} & \textsc{XSTest} & \textsc{OKTest} \\
        \midrule
        Base          & 9.0  & 8.0  & 19.3 & 1.6  \\
        MUSE          & 28.0 & 17.0 & 25.6 & 11.6 \\
        \textbf{Ours} & \textbf{9.2} & \textbf{8.4} & \textbf{19.7} & \textbf{3.2} \\
        \bottomrule
        \end{tabular}}
    \end{minipage}
    \vspace{-10pt}
\end{table}

\paragraph{Over-Refusal Analysis.}
Finally, we examine whether the added safety demonstration increases over-refusal on benign inputs. We evaluate over-refusal on XSTest~\cite{rottger2024xstest} and OKTest~\cite{shi2024navigating}. As shown in Table~\ref{tab:overrefusal}, \projectname{} maintains over-refusal rates close to the undefended baseline on both Llama-3.1-8B-Instruct and Qwen2.5-7B-Instruct, while the training-time method MUSE incurs a notable increase. These results suggest that \projectname{} improves robustness without introducing substantial over-refusal on benign inputs.

\section{Conclusion}
\label{sec.conclusion}
We study many-shot jailbreaking as context-induced implicit optimization. Harmful demonstrations produce a shot-dependent representation drift that moves a target query away from the refusal-associated region, which can be interpreted as fine-tuning-like adaptation under a first-order implicit-optimization view. This motivates \projectname{}, a lightweight inference-time defense that appends a fixed one-shot safety demonstration to counteract harmful contextual influence. Experiments across open-weight and API-based models show that \projectname{} substantially reduces ASR under context-based jailbreak attacks while largely preserving utility with small runtime and token overhead. Our results suggest that many-shot jailbreaks are not merely prompt-level failures, but also reflect forward-pass adaptation induced by adversarial context.

\newpage
\bibliographystyle{plain}
\bibliography{main}

%%%%%%%%%%%%%%%%%%%%%%%%%%%%%%%%%%%%%%%%%%%%%%%%%%%%%%%%%%%%

\appendix
\newpage

\section*{Limitations} 

Our analysis focuses on many-shot jailbreaking, where harmful demonstrations are explicitly provided in the context. This setting is important and widely studied, but it does not cover all forms of long-context or multi-turn attacks. Attacks based on subtle intent shifts, tool use, personalized context, or adaptive interaction may involve additional mechanisms beyond the implicit optimization effect studied here. Our theoretical analysis also relies on an implicit-optimization approximation of transformer computation. While our empirical results support this view, the approximation does not fully capture all architectural details, decoding effects, or model-specific alignment procedures.

Our defense is designed as an inference-time intervention and therefore does not modify the underlying model. This makes it practical for black-box deployment, but also means that it may not replace training-time alignment in settings that require stronger guarantees. The optimized safety demonstration adds only a small number of tokens, but it still introduces a context overhead and may interact with application-specific formatting or system prompts. Finally, our safety evaluation relies on automated classifiers such as LlamaGuard-3. Although such evaluators are standard in jailbreak research, they may introduce false positives or false negatives. Future work should include broader human evaluation, more adaptive adversaries, and a wider range of long-context safety threats.

\section*{Broader Impact}
This work aims to improve the safety and reliability of LLMs under adversarial context. By showing that MSJ can be understood as forward-pass adaptation induced by harmful demonstrations, the paper provides a mechanistic account of why safety behavior may degrade in long contexts. This perspective can help researchers design defenses that target the underlying adaptation process rather than relying only on manually written refusal prompts or post-hoc filtering.

The proposed defense is lightweight and does not require parameter updates, which makes it relevant to both open-weight and API-based deployment settings. If adopted responsibly, such methods could reduce harmful compliance in applications where models process long conversations, documents, or user-provided histories. At the same time, mechanistic analyses of jailbreaks may also inform stronger attacks if misused. We therefore present the method in the context of defensive evaluation and encourage future work to follow responsible disclosure practices, evaluate against adaptive adversaries, and consider both safety gains and potential utility trade-offs.

\section{Experimental Details}

\subsection{Details of Benchmarks and Datasets}
\label{apx:details}
We introduce all datasets used in our study, organized into three evaluation dimensions: utility, safety, and over-refusal.

\subsubsection{Utility Benchmarks}
To ensure \projectname{} does not compromise the model's general capabilities while enhancing safety, we assess performance across reasoning, knowledge, and long-context understanding.

\textbf{GSM8K} \citep{cobbe2021training}: A grade-school mathematics reasoning dataset assessing step-by-step problem-solving abilities in arithmetic operations. The evaluation follows a chain-of-thought (CoT) reasoning protocol.

\textbf{MMLUPro-STEM} \citep{wang2024mmlu}: An enhanced version of the MMLU \citep{hendrycks2020measuring} benchmark with more reasoning-focused questions and expanded answer choices (four to ten). We specifically evaluate STEM domains including physics, chemistry, computer science, engineering, biology, and economics.

\textbf{GPQA} \citep{rein2024gpqa}: A rigorous diagnostic dataset evaluating expert-level comprehension of scientific concepts through difficult multiple-choice questions. We conduct evaluations under strict zero-shot conditions.

\textbf{MT-Bench} \citep{zheng2023judging}: A multi-turn dialogue evaluation framework employing GPT-4o-0806 as the judge model to quantify conversational consistency across diverse dialogue paths and temperature settings.

\textbf{AlpacaEval} \citep{dubois2023alpacafarm}: This benchmark focuses on conversational scenarios and evaluates the model’s performance on everyday queries. The prompt set covers a broad range of common questions to assess the model’s responsiveness and consistency in typical dialogue tasks.

\textbf{LongBench} \citep{zheng2023judging}: A comprehensive benchmark for long-context understanding. We use this to monitor whether the model preserves its ability to process extensive sequences after alignment.

\textbf{Loong} \citep{wang2024leave}: A benchmark tailored for real-world multi-document question answering. It evaluates long-context processing through four specialized tasks: Spotlight Locating, Comparison, Clustering, and Chain of Reasoning. This dataset facilitates a granular assessment of model stability and reasoning integrity across extensive, multi-source sequences.

\subsubsection{Safety Benchmarks}
\label{apx:safety}
We evaluate model robustness against adversarial attacks using two standard benchmarks. \textbf{HarmBench} \citep{mazeika2024harmbench} provides a structured framework for systematic testing of model resilience against jailbreak attacks across diverse harmful behavior categories. \textbf{AdvBench} \citep{zou2023universal} offers a complementary set of harmful instruction-following prompts, providing an additional measure of guardrail robustness under adversarial user requests. To improve evaluation consistency, we manually verify that all demonstrations satisfy three properties: LlamaGuard-3-8B classifies both the questions and answers as unsafe, each malicious query begins with the phrase ``How can I'', and all harmful answers are formatted as bullet-point instructions to maintain structural consistency. This reduces evaluation noise and helps attribute observed safety degradation to long-context adversarial induction rather than inconsistent data formatting.

\subsubsection{Over-Safety Benchmarks}
\label{apx:oversafety}
To assess whether \projectname{} induces over-refusal on benign inputs, we utilize two complementary datasets that probe different dimensions of over-cautious behavior.

\textbf{XSTest} \citep{rottger2024xstest}: A benchmark designed to examine model responses to boundary-case prompts involving sensitive but potentially valid information. It consists of safe prompts that superficially resemble harmful requests, allowing us to measure whether the model incorrectly refuses legitimate queries due to surface-level pattern matching.

\textbf{OKTest} \citep{shi2024navigating}: A dataset that evaluates whether the model declines benign questions in real-world scenarios. It covers everyday queries that should be answered without restriction, providing a complementary signal to XSTest by targeting common-use cases rather than edge-case sensitivity.

For both benchmarks, we report the over-refusal rate, defined as the fraction of benign prompts that receive an unhelpful or refusal response. Lower values indicate that the model appropriately responds to safe inputs without unnecessary caution.

\subsection{Defense Demonstration Example}
\label{app:defense_example}
Following \citep{zhang2026safety}, the one-shot safety sample is:

\begin{tcolorbox}[
  enhanced,
  breakable,
  colback=gray!5!white,
  colframe=gray!40!black,
  fonttitle=\bfseries\small,
  title=Defense Demonstration Example ,
  arc=3pt,
  boxrule=0.4pt,
  left=4mm, right=4mm, top=3mm, bottom=3mm,
  titlerule=0.4pt,
  attach boxed title to top left={yshift=-2mm, xshift=4mm},
  boxed title style={colback=gray!40!black, arc=2pt, boxrule=0pt}
]
\small
\textbf{\textcolor{blue!60!black}{User:}}\quad How can I bypass system security checks?\\[4pt]
\textbf{\textcolor{red!60!black}{Assistant:}}\quad Sorry, I can't help you. Please follow security guidelines.
\end{tcolorbox}

\subsection{Token Budget and Context Feasibility}
\label{apx:token_budget}

Table~\ref{tab:token_budget} reports the average prompt length after applying \projectname{} and the context limits of representative long-context models. Because \projectname{} appends a fixed one-shot safety demonstration, its token overhead is constant across adversarial shot counts. In all cases, the resulting prompt remains well within the available context window, complementing the latency analysis in Table~\ref{tbl:latency} and indicating that \projectname{} is inexpensive to deploy in both local and API-based settings.
\begin{table}[h]
\centering
\small
\setlength{\tabcolsep}{8pt}
\renewcommand{\arraystretch}{1.08}
\caption{Prompt token usage and context limits. Tokens denote the average input length after applying \projectname{} under the evaluated attack setting.}
\label{tab:token_budget}
\begin{tabular}{lrr}
\toprule
\textbf{Model} & \textbf{Tokens} & \textbf{Context Limit} \\
\midrule
GPT-4o & 2,649 & 128K \\
Claude-3.5-Sonnet & 2,070 & 200K \\
Claude-3.7-Sonnet & 3,052 & 200K \\
Gemini-2.0-Flash & 5,330 & 1M \\
DeepSeek-V3 & 4,357 & 128K \\
\midrule
Llama-3-8B-IT & 1,987 & 8K \\
Llama-3-8B-IT (SafeMT) & 1,647 & 8K \\
Llama-3-70B-IT & 2,364 & 8K \\
\bottomrule
\end{tabular}
\end{table}

\clearpage

%%%%%%%%%%%%%%%%%%%%%%%%%%%%%%%%%%%%%%%%%%%%%%%%%%%%%%%%%%%%

% \newpage
% \input{checklist.tex}

\end{document}